# Automated production of batched unclonable micro-patterns anti-counterfeiting labels with strong robustness and rapid recognition speed

Yuzheng He[1], Zunshuai Zhang[1], Yifei Xing[1], Zhiyuan Lang[1], Jinbo Wu[2*], Jiong Yang[1*]

1 Materials Genome Institute, Shanghai Engineering Research Center for Integrated Circuits and Advanced Display Materials, Shanghai University, Shanghai 200444, China.
2 Faculty of Materials Science, Shenzhen MSU-BIT University, Shenzhen, Guangdong, 518172, China.
Emails: jinbowu@smbu.edu.cn; jiongy@t.shu.edu.cn

**Abstract**

Anti-counterfeiting technologies are indeed crucial for information security and protecting product authenticity. Traditional anti-counterfeiting methods have their limitations due to their clonable nature. Exploring new technologies, particularly those based on pixel-level textures is a promising avenue to address the clonable issue due to high encoding capacity. However, research in this field is still in its infancy. This work introduces a new fluorescent anti-counterfeiting label technology with four key characteristics: efficient laser etching, high-throughput fabrication and segmentation, robustness aided by data augmentation, and an exceptionally high recognition speed. To be specific, the etching achieves a speed of 1,200 labels/3s, the high throughput yields a rate of 2,400 labels/4 min, and a total count of 51,966 labels. The number of labels is further augmented to 5,196,600 by implementing arbitrary rotation and brightness variation to enhance the robustness in the recognition procedure. We divide these labels into 44 categories based on differences in patterns. Utilizing machine learning methods, we have achieved a total recognition (including extraction and search process) time per label averaging 421.96 ms without classification, and 40.13 ms with classification. Specifically, the search process with classification is nearly fiftieth times shorter than the non-classification method, reaching 8.52 milliseconds in average. The overall recognition time is much faster than previous works, and achieve an accuracy over 98.7%. This work significantly increases the practicality of pixel-level anti-counterfeiting labels.

**Keywords:** unclonable anti-counterfeiting labels; automated production; augmentation; rapid recognition speed

## Introduction

Under the temptation and drive of economic interests, the forgery of various commodities is increasing day by day, which not only seriously harms the legitimate rights and interests of businesses and consumers, but also threatens national security and human health and so on.[1–3] New and useful anti-counterfeiting materials and technologies have become an urgent need. People have explored various anti-counterfeiting strategies, including inkjet printing[4–7], bar code[7–12], optical anti-counterfeiting[13–16], watermarks[11,12,17,18], laser holograms[18–23] etc. Luminescence anti-counterfeiting as one of the optical anti-counterfeiting technologies, derives from the changeable luminescence behaviors of luminescence materials under the alteration of various external stimuli such as excitation light[24–28], chemical reagent[26,27,29,30], heat[26,27,31], and mechanical force[28,32], luminescence lifetime[33,34] etc. The luminescence anti-counterfeiting has been widely used in many areas such as RMB, food safety, drug packaging, etc.[35–37] Most of this research field focuses on multilevel pattern anti-counterfeiting, which can change more than two times under the regulation of excitation light, luminescence, and so on.[37] The current mainstream research on multilevel fluorescence is to explore various methods for encrypting and decrypting information from fluorescent patterns.[37]

Taking fluorescent anti-counterfeiting as an example, the mentioned multilevel anti-counterfeiting primarily relies on patterns for conveying anti-counterfeiting information, which inherently carries a limited amount of information. In comparison to this approach, texture-based anti-counterfeiting technology operates at the pixel level, as the information used for anti-counterfeiting is small and is achieved through the randomness and uncertainty in the label generation process. This inherent unpredictability grants it natural unclonability. Therefore, texture-based anti-counterfeiting currently holds certain research value in academic community. On the other hand, texture anti-counterfeiting often requires integration with machine learning methods due to the large amount of information that needs to be processed. He et al. designed and manufactured a physically non cloning multimode structured color anti-counterfeiting label based on artificial intelligence (AI) decodable amorphous photonic structures (APS) by introducing an intermediate white layer with water induced transparency between the patterned APS layer and the black background layer.[38] By randomly rotating each of the six initial data to generate 500 images, a total of 3,000 images were obtained to establish the database for deep learning. Samples taken by customers have different exposure, brightness, contrast, or a combination of the above factors were fed into the trained AI for validation. The results showed an accuracy for most of the authentic tags of >0.99 indicating the high precision of the artificial intelligence decoding process. It takes seconds to finish the whole authentication process. Liu et al. develop an inkjet-printable, AI decodable, unclonable, fluorescence security label.[39] About 3,000 images generated by randomly shifting and rotating six original images were provided for AI learning and classification. The deep learning model completed the entire authentication process for tags with different sharpness, brightness, rotation, amplification and mixing of these parameters, and it only takes a few seconds or less, and the matching degree for clear tags reaches about 99 %. Lin et al. developed anti-counterfeiting labels with low reagent cost, simple and fast authentication, huge coding ability and high modifiability.[40] The scale of the labels in this work is smaller than the previous context. In the anti-counterfeiting verification, the two-step method was adopted. The software classified the shape of the patterns of a total of 4,000 images, and then verified the texture. This method improved the efficiency of image recognition. Through the shape matching technology, the overall processing time (smart phone reading +data synchronization +authentication) is 12.17 seconds, and the matching time is 2.69 seconds. 47 authentic images and 27 fake images were tested for verification, achieving accuracy rates of 97.9 % and 100 %, respectively.

There are the current research flaws in pattern and texture anti-counterfeiting. The number of labels is small because of none research on high-throughput synthesis, which is far from practical application. But if the number of labels is increased by a few orders of magnitude, it will also put great pressure on subsequent batch processing of labels, formation of databases and follow-up recognition applications. The recognition speed using the current literature method is slow, and even in databases with only thousands of labels, it also requires a second level recognition time.

This work utilizes laser etching machine, adopts high-throughput preparation, and borrows the latest face recognition technology for etching, batch preparation, and rapid recognition of anti-counterfeiting labels. The etching process can achieve up to 1,200 patterns within a mere 3 seconds and our high-throughput automated shooting speed has reached 2400 labels/batch in 4 minutes. After using image segmentation methods and applying data augmentation techniques as rotating angles and adjusting brightness, the initial set of 51,966 labels has been expanded to 5,196,600 labels. By utilizing feature vector extraction method, a database has been quickly formed. Meanwhile, for a given label to be examined, we can achieve a recognition speed at the level of milliseconds, based on the database with a volume of 4,155,600, which is much faster than the previous literatures. This recognition speed provides rich margin for the recognition of larger data volume databases in the future. Our work has solved the problems of high-throughput etching, batch label generation, and rapid recognition for the practical application of texture anti-counterfeiting labels, making it a significant step forward in industrial applications.

**Results**

**The whole process of anti-counterfeiting label production and recognition.** The schematic diagram of the overall workflow of our work is as figure 1. The entire workflow is segmented into three main stages: etching, label generation, and recognition, where label generation includes detailed steps of high-throughput photography, segmentation, and augmentation. Initially, the prepared thin films undergo etching with numerous pattern arrays. Subsequently, we used the home-made python program to control optical microscope for high-throughput and automatic characterization. After processing the captured photos using a segmentation model, we were able to quickly generate a large number of original labels. The 51,966 original labels were further augmented to 5,196,600 labels by the adjustments of various brightness and rotation angles. For the subsequent recognition process, we ultimately achieved a total time as fast as 40.1 ms and a minimum accuracy of 98.7%. The followings are the details of each step.

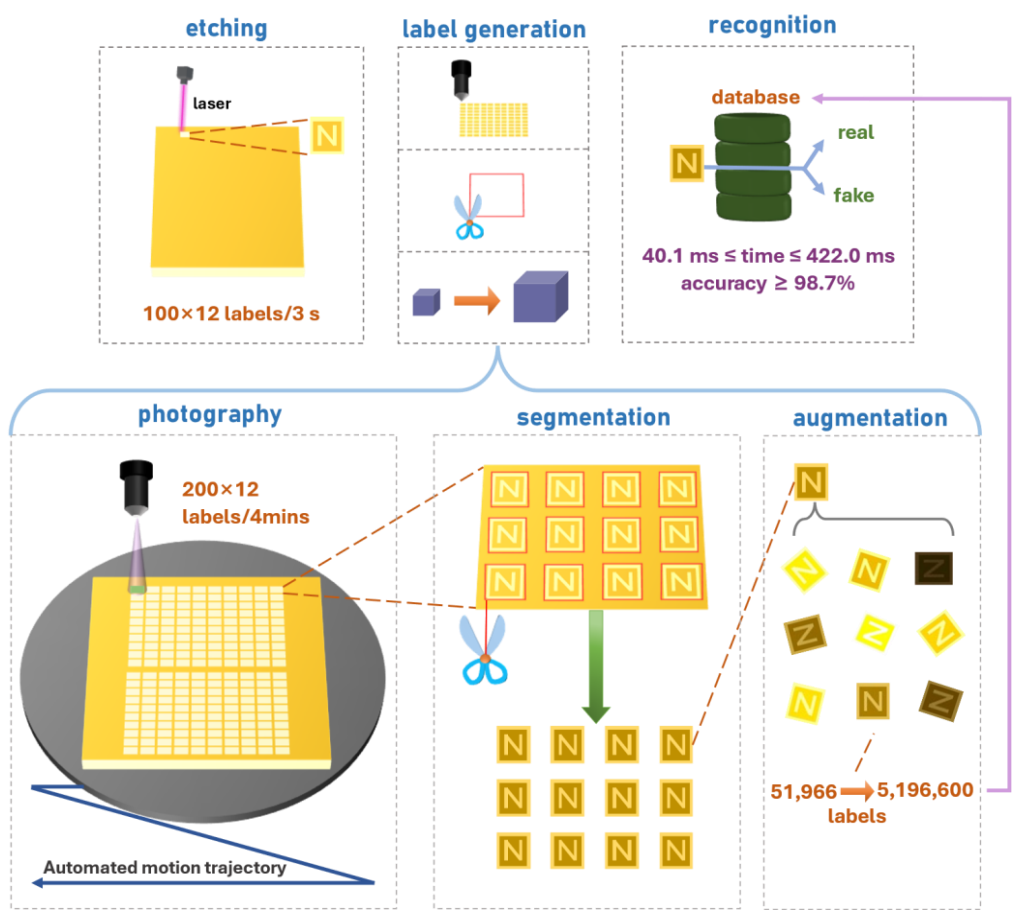

Figure 1. The complete process of etching, label generation, and recognition, with label generation including detailed procedures of high-throughput photography, segmentation, and augmentation.

**Micro-patterns etching quantum dot film.** First, we prepare perovskite quantum dot films and perform high-throughput etching to facilitate subsequent capturing and segmentation. The prepared solution is then spin-coated onto hydrophilic glass substrates treated with Plasma Cleaning Machine. The self-assembly is induced through heating and evaporation, and after constant heating at 75 degrees Celsius for 5 minutes on a hotplate, a perovskite quantum dot nanocrystal film is formed. The formed crystal film has a unique micro-texture that exhibits physical unclonable function characteristics because of the randomness of the crystallization process.[40] After creating the desired pattern array using Computer-Aided Design (CAD), micro-patterns and the borders are realized with high-throughput using a nanosecond laser engraving machine. The etching process achieves a rate where nearly 1,200 labels can be manufactured in about 3 seconds. Our etching is for two purposes: 1) creating a bounding box for segmentation; 2) etching a pattern for subsequent image classification to accelerate recognition speed. It is notable that the output power of the laser is a key parameter to be adjusted. As shown in figure 2a-2d, it can be seen that the larger the etching power, the wider and darker the border. And a distinct border is important for the image segmentation model in the later procedure. Therefore, the power of 128 mW (figure 2d), which gives the most distinct border, was selected. If a higher laser power is employed, although it can make the boundaries clearer, thicker boundaries in the current label array will cause the edges of the labels connected with each other, making it impossible to separate. Therefore, 128 mW is the optimal laser power for this work.

Through CAD and nanosecond laser technology, it is possible to create extremely small and highly controllable labels. This is currently the most suitable technology for mass production in unclonable anti-counterfeiting labels due to the following reasons. Such small size in this work is highly advantageous for generating larger data volumes in the later stages. The second reason is that a greater amount of valuable information is contained in the current labels. It is evident as shown in figure 2d that the vast majority of the entire label consists of anti-counterfeiting texture information with an average content of about 75%, whereas previous work using brushing methods relied solely on the texture information applied through brushing.[40] The third reason is a significant improvement in the process of fabrication. The procedure in our work, including utilizing a spin coater for uniform coating and employing a laser engraving machine for automated etching, greatly reduces the need for manual labor, eliminating the requirement for a homemade coating device by hand.[40] The fourth reason is that etching process achieves a fast rate where 1,200 labels can be manufactured in just 3 seconds. Furthermore, our method presents clear boundaries. The clarity of the boundaries obtained by laser engraving is significantly superior to those achieved by etching and subsequent brushing.[40] The scanning electron microscope (SEM) images are shown in figure 2e-2g. By comparing figure g1 and g2, it can be seen that the grains in the etched area are more uniform and smaller than those in the non etched area.

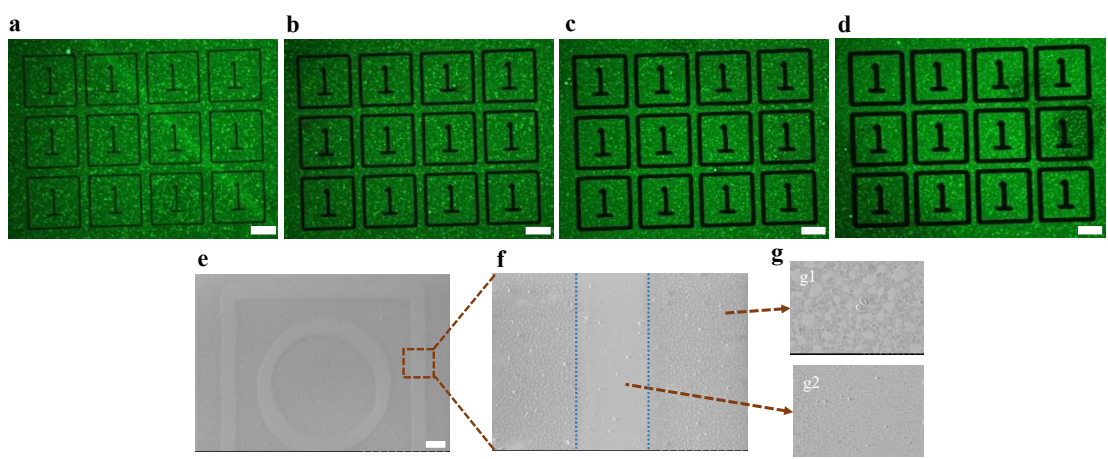

Figure 2. a, b, c, d) The fluorescence image of the generated micro-patterns film at 17, 38, 73, 128 mW etching power (There is no obvious change in the etching area at 7 mW). The scale bar is 100 μm. e) SEM images of a perovskite quantum dot film after etching micro patterns. The scale bar is 100 μm. f) SEM images of etching lines of micro patterns. The scale bar is 20 μm. g) SEM images of part of film. (g1) is the enlarged view of the non etched area. (g2) is the enlarged view of the etched area.

**High throughput photography and label segmentation methods.** This is the crucial step in achieving batch experiments, and our automated photography method is as follows. As shown in figure 1, during the process of high-throughput photography, in one batch we have etched 200 images, each one containing 12 labels. We imported the coordinates of the 200 images into the microscope's automatic photography module, enabling the automated capturing. These coordinates were generated from the array arrangement designed during the CAD drawing, together with the physical location of the first image in the field of the microscopy. A home-made Python program is utilized to generate the file containing all the coordinates simultaneously. Through the position information, the microscope automatically took 200 images; the number touches the uplimit of the instrument. As mentioned, each image is equipped with 12 labels, and this is attributed to two reasons: firstly, to maximize efficiency by fully utilizing the 1,920×1,440 resolution; secondly, to prevent focusing issues. Through testing, having too many labels in one image can make focusing more challenging on the current microscopy, especially due to the unevenness of the samples. This method lays the foundation for automated production of anti-counterfeiting labels.

Due to the fact that each image (containing 12 labels) is less than $1mm^2$, it is inevitable that the labels are not absolutely aligned within the image during the capture process, as shown in figure 2 and figure 3; a proper label segmentation method is needed to fulfill the automation workflow. Automated image segmentation was unnecessary in previous work, due to small number of labels[38–40]. However, it has become crucial in our high-throughput task. Additionally, given the relatively clear square contours, we can achieve the label segmentation effectively. This method processes images in four steps. First, the image is converted into a grayscale image. Second, the image is smoothed using a low-pass filter and then subjected to binarization to remove noise. Next, all contours are found, and the largest 12 contours are selected based on the number of labels in one image. Lastly, the minimum bounding rectangle for each contour is calculated, the original image is cropped accordingly, and the labeled images are stored as needed. We process all the images by the label segmentation method. Since there are 12 labels in each image and 200 images in each batch, about 2,400 labels were obtained after the label segmentation in about 40 seconds. Including the 4 minutes of shooting, the overall high-throughput process of characterization takes about 5 minutes per batch, as shown in figure 3. After repeating this process for about 20 times, we finally obtained 51,966 original labels.

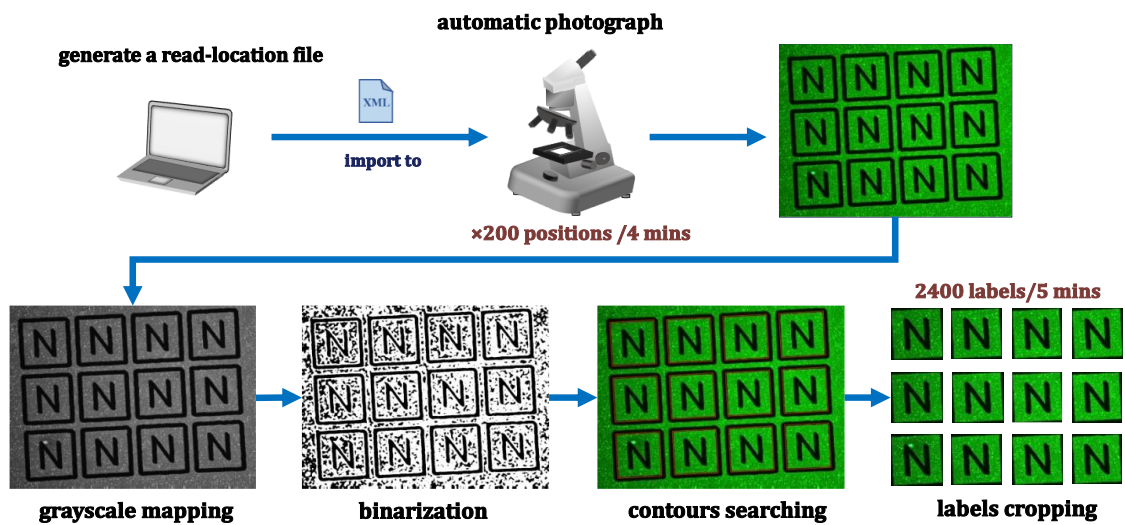

Figure 3. Flow chart of high-throughput photograph and label segmentation

**Data augmentation and classification.** In order to boost the robustness of the machine learning model, we perform data augmentation. In practical applications, the image quality of anti-counterfeit labels may be influenced by the factors such as different lighting conditions and shooting angles. The significance of data augmentation lies in the substantial increase of the quantity of training data, and the robustness of the recognition process. By applying random rotations and brightness adjustments to a single label, we generate a total of 5,196,600 labels, i.e., 100 augmented ones per label. The rotation angle varies randomly from 0 to 360 degrees. The colors of the labels could be represented in the HSV color space, where V (Value, ranging from 0 to 255) represents the brightness. The brightness adjustment in the data augmentation process involves adding the V component by an 'x' value where 'x' $\in$ (-45, 90). The label examples after rotation angle and brightness adjustments (under a fluorescent microscope) are as illustrated in figure 4(a). The histogram in figure 4(b) illustrates the brightness variations in the example shown in figure 4(a). It can be observed that the peak of the V-values for labels shifts towards higher values after brightening, while it shifts towards lower values after dimming the brightness.

The random and uncontrollable crystallization process will also cause the formed crystal film to have a unique texture[40]. In order to make the subsequent classification more convenient and rapid as the previous work[40], we etched several different types of patterns, including drawing, digit, letter and Chinese character, as shown in figure S6. And there are about 44 groups of patterns in total. These patterns show the richness of etched patterns and a variety of patterns can be etched without being limited by whether the patterns are solid or not[40], and can classify 5,196,600 labels into 44 groups of texture database with different patterns. In the following recognition process, we selected 4,155,600 labels from the total 5,196,600 ones as the database (marked as real), with the rest 1,041,000 labels used as the fake ones (outside the database). We further explore the impact from the label classification on the recognition speed.

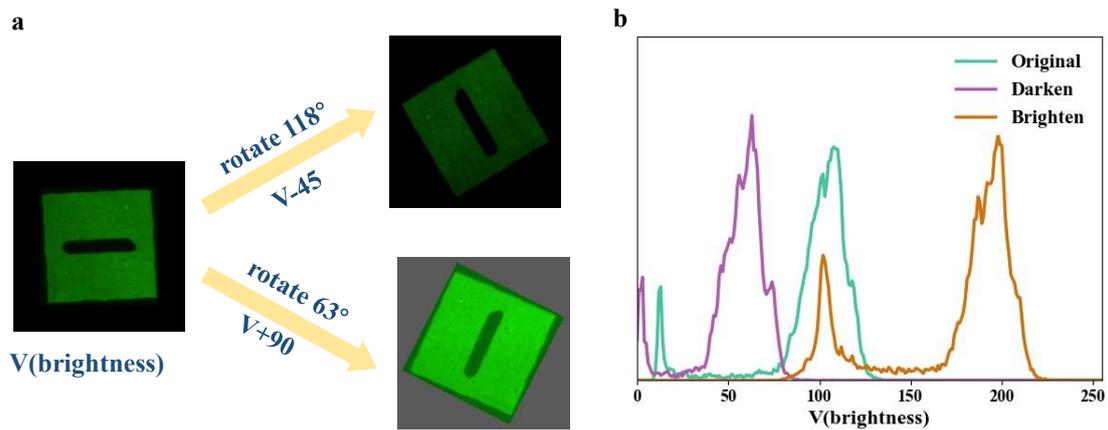

Figure 4. (a) Examples after rotation angle and brightness adjustments (under a fluorescent microscope) and

(b)Histogram of corresponding Brightness (V)

**Overall recognition process.** These are two steps in the recognition process, feature vector extraction and search in the database. For feature vector extraction, we utilized the ResNet-50 network structure of the convolutional neural network.[41] The loss function of this model is the Additive Angular Margin Loss (ArcFace), which significantly enhances discriminative power.[42] And the method here is borrowed from a popular line of research in the field of face recognition, which involves adopting margins in the well-established softmax loss function to maximize class separability. The feature vectors of above-mentioned 4,155,600 labels have been extracted and stored into the database. For the search step of one particular label (inside or outside the database), we firstly augmented the model with a layer of multilayer perceptron, used for the classification of patterns if needed. By employed the Facebook AI Similarity Search (Faiss) retrieval framework, we were able to efficiently search the whole database to verify if the match is inside or not.

We selected 10,000 real and fake labels to test the recognition speed and accuracy. The frequency distribution histogram is shown in figure 5. We estimate based on the average value in the statistics. Figure 5a demonstrate that for real labels without pattern classification, the feature vector extraction time is approximately 30.70 ms per label in average, and the search time in the database is about 387.06 ms. Figure 5b shows that, without classification, the feature extraction time for fake labels is 30.93 ms, and the search time is 395.23 ms. When pattern classification is applied, as illustrated in figure 5c and 5d, both real and fake labels show a slight increase in feature vector extraction time, i.e., 31.29 ms for real and 31.73 ms for fake. However, the search time significantly drops to 8.34 ms for real labels and 8.69 ms for fake labels, which is almost one fiftieth of the search time for labels without classification. Therefore, for the unclassified method, the total times used for real and fake labels are respectively 417.76 ms and 426.16 ms. With classification, these times reduce to 39.83 ms for real and 40.42 ms for fake labels, indicating a significant efficiency gain. In terms of accuracy, a relatively high level was achieved. For the unclassified method, the accuracy in identifying real and fake labels was 99.77% and 98.7%, respectively. With the application of the classification method, these accuracies were 98.87% and 98.73%, respectively.

The demonstrated capability to process the recognition under a vast database (containing 4,155,600 labels) within millisecond level shows exceptional efficiency. Such rapid processing is vital in practical scenarios where there's a need to swiftly handle a large number of labels. Furthermore, achieving at least 98.7% accuracy rate in recognition is particularly impressive given the volume of the database. This high level of accuracy is crucial in anti-counterfeiting efforts, ensuring reliability and trust in the methods. Previous works[38-40] typically used data volumes of about 3,000 to 4,000, also achieving accuracy rates close to 98%, with processing times generally in the seconds. Despite nearly similar accuracy rates, our work demonstrates a magnitude of improvement in both time and data volume compared to previous efforts. The implications of these findings are significant for future practical applications. They pave the way for the advanced recognition of unclonable anti-counterfeiting labels on a much larger scale, making the technology more applicable and relevant in real-world scenarios. This advancement not only enhances the capability of anti-counterfeiting technologies but also offers valuable insights and benchmarks for future developments in this field.

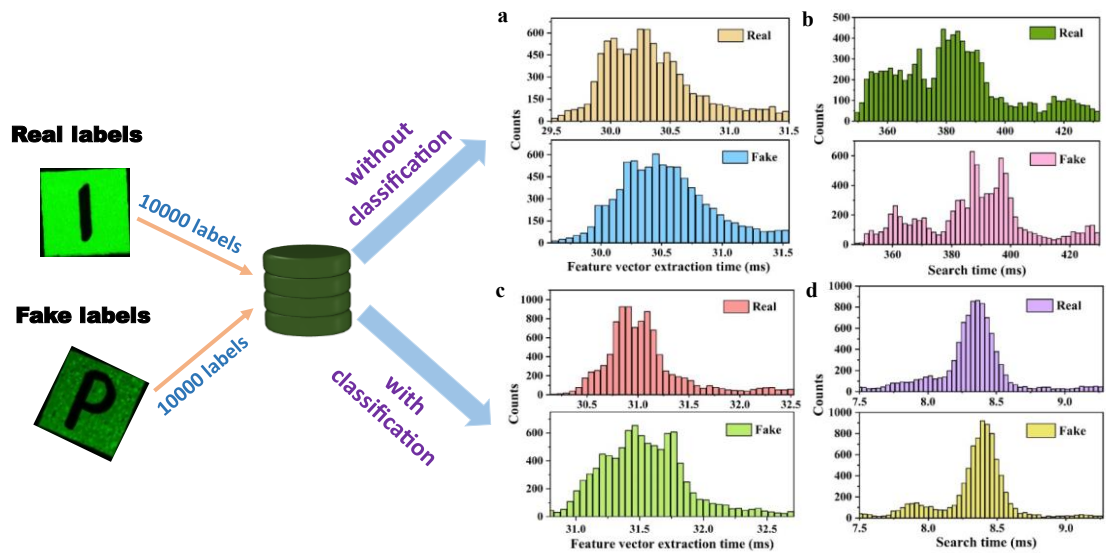

Figure 5. The frequency distribution histogram of (a) the feature vector extraction time and (b) search time for 10000 real or fake labels with pattern classification and (c) the feature vector extraction time and (d) search time for 10000 real or fake labels with pattern classification.

**Discussion**

Since we are conducting pixel-level recognition and matching, reducing the etched area implies that the textured region will be larger. This increases the number of pixels we can identify, leading to improved accuracy. Meanwhile, the time used for etching could also be saved. In fact, we can even create labels without patterns to maximize the effective information content of each one, and still achieve an acceptable speed given the current volume of labels. The type of label without a pattern, as illustrated in figure

6, offers the advantage of eliminating the need for CAD drawing, etching, and segmentation; it only needs cutting based on size as indicated by the dashed line in the figure 6. This results in maximizing the effective information. However, the drawback is that, due to the absence of classification, the identification process takes a slightly longer time as discussed above. In practical applications, cutting along virtual lines is required.

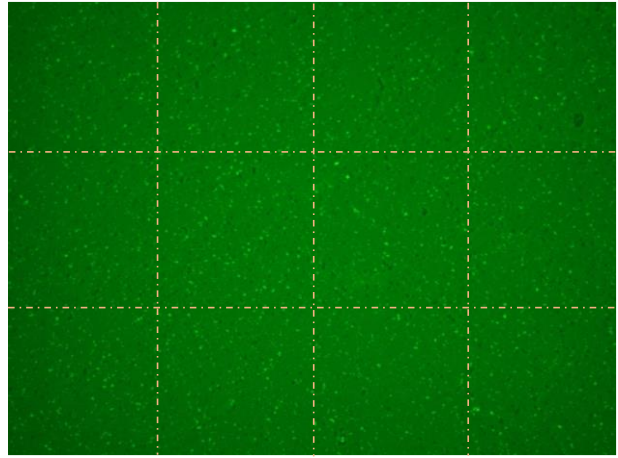

Figure 6. The schematic diagram depicting the virtual cutting of patternless labels

In summary, by optimizing the laser etching parameters, we obtain perovskite quantum dots nanocrystalline micro-patterns with micro size, high resolution, and different types, and the process achieves a rapid speed where 1,200 labels are produced in just 3 seconds. By designing a pattern array and inputting it into a python program to obtain files that can be read by the microscope automatic photographing module, we have achieved high throughput photography, enabling 2,400 anti-counterfeiting labels to be shot within about 4 minutes. The image segmentation method divides these images into final anti-counterfeiting labels. And each label has a unique micro texture. We generate a total of 5,196,600 labels by changing rotations and brightness randomly to a single label. By applying the ResNet-50 network for feature vector extraction and Faiss for search, the recognition process under millions of labels demonstrates strong advantages over literatures in terms of efficiency (around 40 ms per label) and accuracy ($\geq$ 98.7%), especially with classifications of patterns. For future data that may reach tens of millions or even billions in scale, our feature vector extraction time can still be maintained at around 30 milliseconds. And the pattern classification and / or a more efficient retrieval model are also helpful to improve the overall efficiency in the future.

**Methods**

**Preparation of micro-patterns of perovskite quantum dot film.** In our method, We take 0.1468g of $PbBr_2$ (0.4 moles) and 0.0852g of CsBr (0.4 moles), along with 1 ml of oleic acid and 0.5ml of oleylamine, and thoroughly mix them with 10 ml of dimethylformamide. We use ultrasound agitation until the mixture becomes clear, resulting in the precursor solution for perovskite formation. Next, we heat 10 ml of toluene on a heating plate at 35 degrees Celsius for 10-15 minutes. Afterward, we carefully add 1 ml of the precursor solution to the toluene. We then stir the mixture thoroughly using a magnetic stirrer under constant temperature for 5 minutes to ensure uniform mixing. After mixing the obtained crude solution with ethyl acetate 1:2, we centrifuged it at 9,000 rpm for 3 minutes, removed the precipitate, and dispersed it into 2 ml of n-hexane to obtain yellow-green perovskite quantum dot solution. The prepared perovskite quantum dot solution was spin-coated onto the hydrophilic substrate after PLASMA treatment, and self-assembly was induced by heating and evaporation. After a heat process of 5 minutes on a 75-degree heating plate, the perovskite nanocrystalline film was crystallized.

**Synthesis of perovskite quantum dot nanocrystal film.** Figure S1, Supporting Information, shows the X-ray diffraction (XRD) characterization data of the resulting sample. The main diffraction peaks at 15.0° (100), 21.2° (110), 30.2° (200) and 43.4° (220) were all characteristic peaks of $CsPbBr_3$ quantum dot, indicating that it is a monoclinic phase. In addition, the fluorescence spectrum (figure S2 and S3, Supporting Information) showed that $CsPbBr_3$ quantum dot film had an absorption peak at 334 nm, a photoluminescence (PL) peak at 513 nm, and a stokes shift from 334 to 513 nm between emission and band-edge absorption. Furthermore, it can be seen from figure S4, Supporting Information, that the short lifetime $\tau 1$ of the prepared $CsPbBr_3$ quantum dot film was 10.0 ns, the long lifetime $\tau 2$ was 53.3 ns, the average lifetime was 44.9 ns, and the corresponding PL quantum yield (PLQY) was 89.4%. The high-resolution transmission electron microscopy image shows as figure S5 that the lattice fringes of perovskite quantum dots is 0.296 nm.

**The specific details of image segmentation methods.** This method is divided into four steps in total. The first step is to convert the image into a grayscale image and use the Sobel operator to calculate the gradients in the x and y directions. Then, we subtract the gradients in the y direction from the x direction. Through this subtraction, we obtain an image region with high horizontal gradients and low vertical gradients. The second step is to remove noise from the image, which includes two parts. The first part is to use a low-pass filter to smooth the image (9 x 9 cores), which will help smooth out high-frequency noise in the image. The goal of a low-pass filter is to reduce the rate of change of the image. For example, replacing each pixel with the mean of the surrounding pixels can smooth and replace areas with significant intensity changes. The second part is binarization of blurred images. Any pixel in the gradient image that is not greater than 90 is set to 0 (black). Otherwise, pixels are set to 255 (white). The third step is to find all the contours of the image. Specifically, we use cv2.findContours() function to search for all contours, then calculate the area of all contours found, and finally select the top K contours based on the number of labels K in the image (K=12). For contour detection model, we used the ResNet-18 network structure of convolution neural network. The fourth step is to calculate the minimum circumscribed matrix corresponding to the contour using the cv2.minAreaRect function, then crop the original image based on the coordinates provided by the circumscribed matrix, and finally store the cropped labels according to the contour or labels as needed.

**The network architecture in the recogniton process.** The network architecture of ResNet-50 is shown in Figure 7. This network is used to extract feature vectors from labels and search for similar vectors in a vector database to determine authenticity. Below the Bottleneck module in Figure 7, the changes in the number of channels in the model are displayed. The specific structure of the Bottleneck is shown in Figure 8.

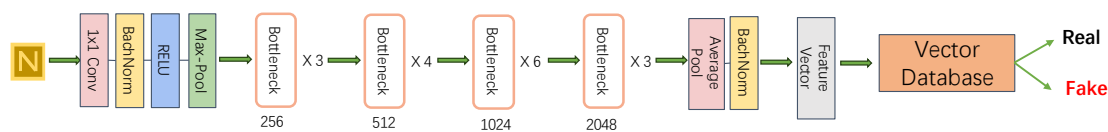

Figure 7. The network architecture during the identification process

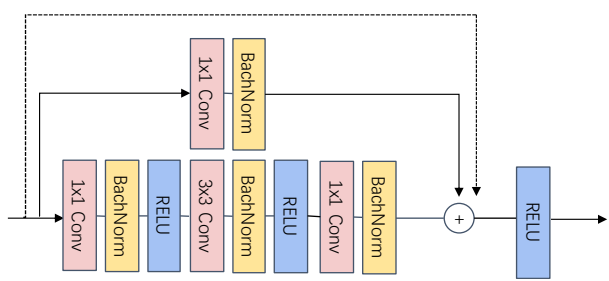

Figure 8. The specific structure of the Bottleneck